\documentclass[aps,prb,superscriptaddress,10pt]{revtex4-2}
\usepackage[utf8]{inputenc}
\usepackage[T1]{fontenc}
\usepackage{threeparttable}
\usepackage{booktabs}
\usepackage{float}
\usepackage{multirow}
\usepackage{graphicx}
\usepackage{dcolumn}
\usepackage{bm}
\usepackage{bookmark}
\usepackage{color}
\usepackage{xcolor}
\usepackage{amsmath,amssymb,bm}
\usepackage{threeparttable}
\usepackage{makecell}
\begin{document}

\title{TorchNEP: Ultra-Efficient and Accurate Training of Neuroevolution Potentials
}

\author{Yong-Chao Wu}
\email{yongchao.wu@aalto.fi}
\affiliation{CSM group, Department of Applied Physics, Aalto University, P.O. Box 15600, 00076 Aalto, Espoo, Finland}

\author{Xiaoya Chang}
\affiliation{MSP group, Department of Applied Physics, Aalto University, P.O. Box 15600, 00076 Aalto, Espoo, Finland}

\author{Tero M\"{a}kinen}
\affiliation{CSM group, Department of Applied Physics, Aalto University, P.O. Box 15600, 00076 Aalto, Espoo, Finland}

\author{Amin Esfandiarpour}
\affiliation{NOMATEN Centre of Excellence, National Center for Nuclear Research, A. Soltana 7 St., 05-400 Otwock, Poland}

\author{Jian-Li Shao}
\affiliation{State Key Laboratory of Explosion Science and Safety Protection, Beijing Institute of Technology, Beijing, 100081, China}

\author{Tapio Ala-Nissila}
\affiliation{MSP group, Department of Applied Physics, Aalto University, P.O. Box 15600, 00076 Aalto, Espoo, Finland}
\affiliation{{Interdisciplinary Centre for Mathematical Modelling and Department of Mathematical Sciences, Loughborough University, Loughborough, Leicestershire LE11 3TU, United Kingdom}}

\author{Zheyong Fan}
\email{brucenju@gmail.com}
\affiliation{College of Physical Science and Technology, Bohai University, Jinzhou, China}

\author{Mikko Alava}
\email{mikko.alava@aalto.fi}
\affiliation{CSM group, Department of Applied Physics, Aalto University, P.O. Box 15600, 00076 Aalto, Espoo, Finland}

\date{\today}

\begin{abstract}
Neuroevolution Potential (NEP) is one of the most efficient machine-learned interatomic potential frameworks for large-scale atomistic simulations. However, its original training strategy remains computationally demanding, limiting systematic exploration of model architectures and training protocols. Here, we present TorchNEP, a PyTorch-based implementation of NEP that combines analytically derived gradients, adaptive optimization, and a two-stage training strategy. TorchNEP accelerates training by more than two orders of magnitude while maintaining full compatibility with existing NEP models. We further show that the improvement in predictive accuracy primarily originates from the two-stage training protocol rather than the optimization algorithm itself. Across diverse benchmark datasets, TorchNEP consistently improves force and stress predictions while maintaining comparable or improved energy accuracy. Benchmark evaluations on elemental and alloy systems demonstrate enhanced predictive performance for both atomic configurations and key materials properties. Furthermore, we show that increasing model complexity does not necessarily improve predictive performance despite reducing training errors. Overall, TorchNEP provides an efficient and flexible training framework for developing more accurate and robust machine-learned interatomic potentials.

\end{abstract}

\maketitle

\section{Introduction}

Machine-learned interatomic potentials (MLIPs) have transformed atomistic simulations by bridging the accuracy of first-principles calculations and the length and time scales accessible to molecular dynamics (MD) simulations~\cite{mishin2021machine}. By learning the potential-energy surface directly from electronic-structure data, MLIPs enable simulations with near density-functional-theory (DFT) accuracy at a computational cost several orders of magnitude lower than direct quantum-mechanical calculations. Over the past decade, numerous MLIP frameworks have been developed, including Gaussian approximation potential (GAP)~\cite{sascha2023gaussian}, moment tensor potential (MTP)~\cite{novikov2022magnetic,evgeny2023mlip3}, atomic cluster expansion (ACE)~\cite{ralf2019atomic}, MACE~\cite{batatia2022mace}, deep potential (DP)~\cite{zeng2023deepmd}, and neuroevolution potential (NEP)~\cite{fan2021neuroevolution}. Among these approaches, NEP occupies a unique position by combining compact physically motivated descriptors with a lightweight neural-network architecture, thereby achieving an exceptional balance between predictive accuracy and computational efficiency. Implemented within the GPUMD package~\cite{gpumd}, NEP has been successfully applied to a broad range of materials, including elemental solids, low-dimensional materials, interfaces, alloys, liquids, and complex disordered systems~\cite{ying2025advances}. More importantly, its highly optimized implementation enables simulations at scales that remain challenging for many other MLIP frameworks. Recent work by Song \textit{et al.}~\cite{song2024general} demonstrated NEP-based MD simulations of metallic systems containing more than $10^8$ atoms using only eight NVIDIA A100 GPUs, highlighting its potential as a practical tool for extreme-scale atomistic simulations.

As MLIPs continue to grow in complexity and scope, the demand for efficient model training has become increasingly important. Modern MLIP development is no longer limited to fitting a single potential~\cite{batatia2025afoundation}. Instead, it increasingly relies on iterative workflows such as active learning~\cite{evgeny2023mlip3}, committee-model uncertainty estimation~\cite{zhang2020depgen}, automated dataset expansion~\cite{chen2025neptrain}, and systematic exploration of model architectures and hyperparameters~\cite{akiba2019optuna}. These tasks often require training tens to hundreds of candidate models, making training efficiency a critical factor in practical MLIP development. Consequently, the computational cost of model construction is emerging as a serious bottleneck, particularly for large and chemically complex datasets. While NEP has achieved remarkable success as an efficient simulation engine, comparatively little attention has been devoted to improving the efficiency of its training workflow. In addition, the effects of optimization algorithms, training protocols, and model complexity influence predictive performance remain insufficiently explored.

Most modern MLIP frameworks, including DP~\cite{zeng2023deepmd}, NequIP~\cite{batzner2022e3}, and MACE~\cite{batatia2022mace}, employ gradient-based optimization together with mini-batch training, adaptive learning-rate scheduling, and dynamically adjusted loss weights. In contrast, the original NEP implementation relies on the separable natural evolution strategy (SNES)~\cite{schaul2011high}, a gradient-free population-based optimizer. Although SNES is robust and has played an important role in the success of NEP, it generally requires a large number of model evaluations and can become computationally expensive when repeatedly training models during active-learning cycles or hyperparameter exploration. Recently, preliminary studies have explored gradient-based optimization for NEP training~\cite{huang2026efficient}, suggesting a possible alternative to evolutionary optimization. However, it remains unclear whether gradient-based approaches can fully exploit the NEP framework and, more importantly, how optimization algorithms and training protocols affect both computational efficiency and predictive performance. Addressing this question is particularly important because improvements in training efficiency do not necessarily translate into better predictive accuracy or transferability.

In this work, we present TorchNEP, a PyTorch-based implementation of the NEP framework that preserves the original descriptor formalism and neural-network architecture while replacing evolutionary optimization with analytical gradients and the Adam optimizer. 
While TorchNEP adopts gradient-based training in place of the original evolutionary algorithm, we preserve the ``NEP'' suffix to maintain continuity with our prior work and to acknowledge the conceptual origin of the potential form.
We benchmark TorchNEP on a diverse collection of datasets spanning elemental systems, molecular and condensed-phase materials, interfaces, and multicomponent alloys. Compared with the original NEP implementation, TorchNEP achieves more than two orders of magnitude acceleration in training while maintaining full compatibility with existing NEP models. We further investigate the respective roles of optimization algorithms, training protocols, and model complexity through extensive benchmarks on multiple datasets and a comprehensive 16-element test suite. Together, these studies provide both an efficient training framework and practical insights into the development of accurate and robust NEP models.

\section{Results}

\subsection{TorchNEP framework and optimization strategy}

TorchNEP implements the neuroevolution potential (NEP) framework within a gradient-based training scheme. Following the original NEP formulation~\cite{fan2021neuroevolution,fan2022gpumd,song2024general}, the total potential energy of a structure containing $N$ atoms is decomposed into atomic contributions,
\begin{equation}
  E = \sum_{i=1}^{N} E_i ,
\end{equation}
where each atomic energy $E_i$ is predicted from a local descriptor vector $\bm{q}^{i}$ that encodes the atomic environment of atom $i$ within finite radial and angular cutoff radii. Before entering the neural network, each descriptor component is normalized using the corresponding minimum and maximum values evaluated over the training set,
\begin{equation}
  \tilde{q}^{\,i}_{\nu}
  =
  \frac{
    q^{i}_{\nu} 
  }{
    q^{\max}_{\nu} - q^{\min}_{\nu}
  } ,
  \label{eq:norm}
\end{equation}
where $\nu$ denotes the descriptor component. These normalization factors are computed once before training and kept fixed throughout optimization.

The atomic energy is represented by a single-hidden-layer feed-forward neural network,
\begin{equation}
  E_i
  =
  \sum_{\mu=1}^{N_{\mathrm{neu}}}
  w^{(1)}_{t_i,\mu}
  \tanh
  \left[
  \sum_{\nu=1}^{N_{\mathrm{des}}}
  w^{(0)}_{t_i,\mu\nu}
  \tilde{q}^{\,i}_{\nu}
  -
  b^{(0)}_{t_i,\mu}
  \right]
  -
  b^{(1)} ,
  \label{eq:ann}
\end{equation}
where $t_i$ is the chemical species of atom $i$, $N_{\mathrm{neu}}$ is the number of hidden neurons, and $N_{\mathrm{des}}$ is the descriptor dimension. The weights $w^{(0)}_{t_i}$ and $w^{(1)}_{t_i}$, as well as the hidden-layer bias $b^{(0)}_{t_i}$, are species dependent, whereas the output bias $b^{(1)}$ is shared by all species. The descriptor vector $\bm{q}^{i}$ contains radial and angular components constructed from Chebyshev radial basis functions and spherical harmonics, and more details can be found in the Methods section.

The training objective combines energy, force, and virial losses,
\begin{equation}
  \mathcal{L}
  =
  \lambda_E \Delta_E
  +
  \lambda_F \Delta_F
  +
  \lambda_V \Delta_V ,
  \label{eq:loss}
\end{equation}
with
\begin{align}
  \Delta_E
  &=
  \frac{1}{N_{s}}
  \sum_s
  \left(
  \frac{\hat{E}_s-E_s}{N_{\mathrm{a}}^{s}}
  \right)^2 ,
  \\
  \Delta_F
  &=
  \frac{1}{3N_{\mathrm{a}}}
  \sum_i
  \left\|
  \hat{\bm{F}}_i-\bm{F}_i
  \right\|^2 ,
  \\
  \Delta_V
  &=
  \frac{1}{9N_{s}}
  \sum_s
  \left\|
  \frac{\hat{\bm{W}}_s-\bm{W}_s}{N_{\mathrm{a}}^{s}}
  \right\|^2 .
\end{align}
Here, hats denote model predictions, $N_{s}$ and $N_{\mathrm{a}}$ are the number of structures and the total number of atoms in the batch, respectively. $N_{\mathrm{a}}^{s}$ is the number of atoms in structure $s$.

\begin{figure}[H]
    \centering
    \includegraphics[width=1\linewidth]{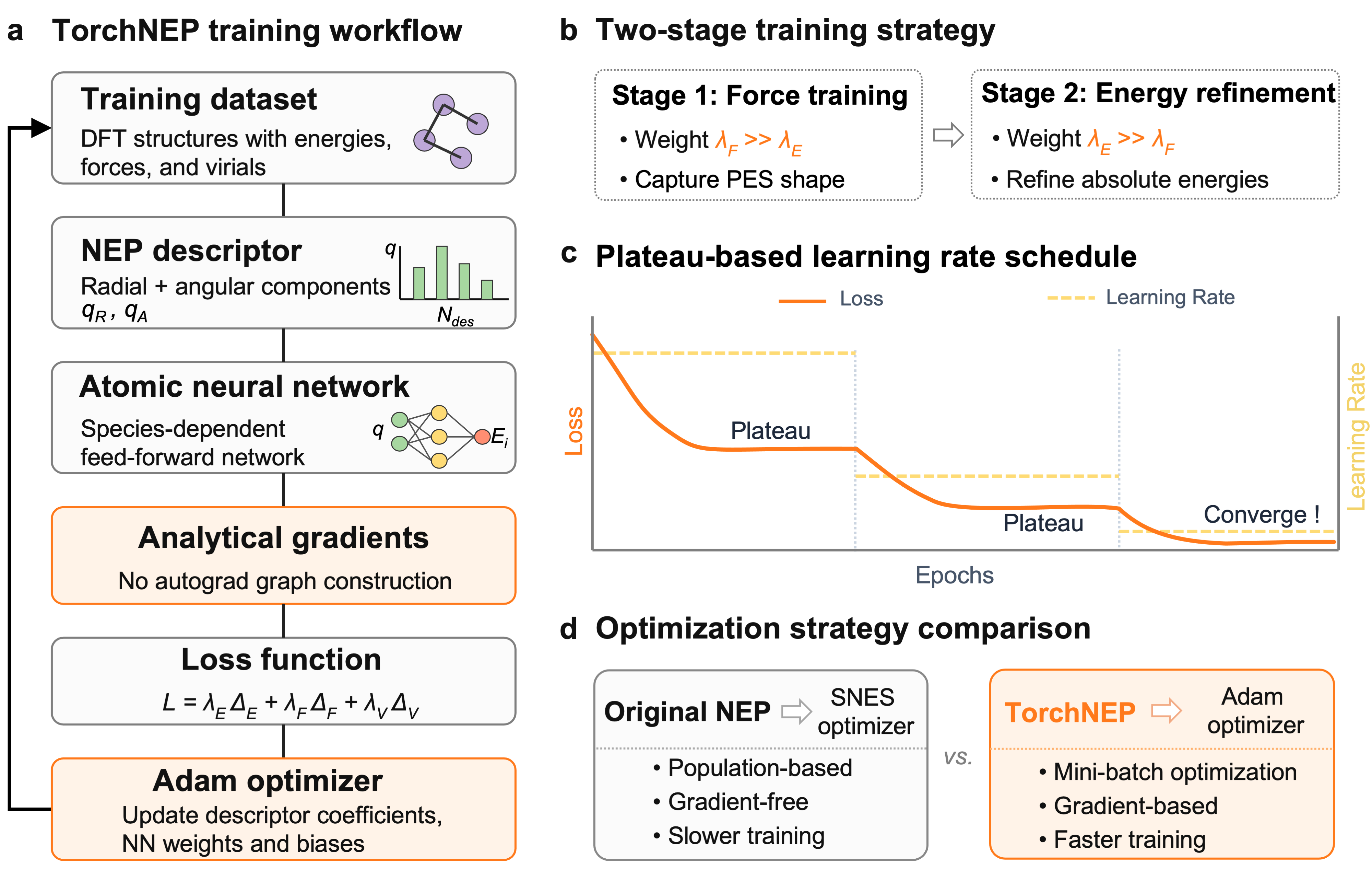}
    \caption{\textbf{TorchNEP training strategy.} (a) Overall TorchNEP training workflow. (b) Two-stage training strategy consisting of force-prioritized training and energy refinement stages. (c) Plateau-based learning-rate decay schedule used during optimization. (d) Comparison of the optimization strategies employed in the original NEP and TorchNEP frameworks. Components introduced in TorchNEP are highlighted in orange.}
    \label{fig:workflow}
\end{figure}

TorchNEP directly optimizes the parameter vector $\bm{\theta}$, including both descriptor coefficients and neural-network parameters, using analytical gradients and the Adam optimizer~\cite{kingma2015adam} with the AMSGrad variant \cite{reddi2018onthe}, which
adapts the per-parameter learning rate from the first- and second-order moments
of the gradient. At training step $t$, that is, the $t$-th parameter update on a mini-batch, the gradient reads
\begin{equation}
  \mathbf{g}_t = \nabla_{\boldsymbol{\theta}}
  \mathcal{L}(\boldsymbol{\theta}_{t-1}).
\end{equation}
 The biased first- and
second-moment estimates are updated as
\begin{align}
  \mathbf{m}_t &= \beta_1 \mathbf{m}_{t-1} + (1-\beta_1)\,\mathbf{g}_t, \\
  \mathbf{v}_t &= \beta_2 \mathbf{v}_{t-1} + (1-\beta_2)\,\mathbf{g}_t^{2},
\end{align}
where $\beta_1$ and $\beta_2$ are the exponential decay rates for the first and
second moments, and $\mathbf{g}_t^{2}$ denotes the element-wise square. After
bias correction,
\begin{equation}
  \hat{\mathbf{m}}_t = \frac{\mathbf{m}_t}{1-\beta_1^{\,t}}, \qquad
  \hat{\mathbf{v}}_t = \frac{\mathbf{v}_t}{1-\beta_2^{\,t}},
\end{equation}
where the superscript $t$ on $\beta_1^{\,t}$ and $\beta_2^{\,t}$ denotes the
$t$-th power, the AMSGrad variant maintains the running maximum of the second moment,
\begin{equation}
  \hat{\mathbf{v}}_t^{\max} = \max\!\left(\hat{\mathbf{v}}_{t-1}^{\max},\,
  \hat{\mathbf{v}}_t\right),
\end{equation}
and updates the parameters according to
\begin{equation}
  \boldsymbol{\theta}_t = \boldsymbol{\theta}_{t-1}
  - \eta_t \, \frac{\hat{\mathbf{m}}_t}
  {\sqrt{\hat{\mathbf{v}}_t^{\max}} + \epsilon},
\end{equation}
where $\eta_t$ is the learning rate at step $t$ and $\epsilon$ is a small
constant for numerical stability. We use the default values $\beta_1 = 0.9$,
$\beta_2 = 0.999$, and $\epsilon = 10^{-8}$. The overall TorchNEP training workflow is illustrated in Fig.~\ref{fig:workflow}a. To improve optimization efficiency, we employ a two-stage training strategy, as shown in Fig.~\ref{fig:workflow}b. In Stage~1, a relatively large force weight $\lambda_F$ is used to prioritize force accuracy and capture the local shape of the potential-energy surface. In Stage~2, the energy weight $\lambda_E$ is increased while the force weight $\lambda_F$ and the learning rate are reduced, allowing the model to refine absolute energies without sacrificing force accuracy.
Within each stage, the learning rate $\eta_t$ is adjusted dynamically using a plateau-based scheduler (\texttt{ReduceLROnPlateau}; Fig.~\ref{fig:workflow}c) that operates at the epoch level: it is held fixed across all optimization steps within an epoch and updated only at epoch boundaries. Denoting by $\eta_e$ the learning rate used throughout epoch $e$, the epoch-averaged training loss $\mathcal{L}$ is monitored, and the learning rate is reduced by a factor $\gamma \in (0,1)$ whenever this loss fails to improve for $p$ consecutive epochs:
\begin{equation}
\eta_{e+1}=
\max\!\left(
\gamma\, \eta_e,\,
\eta_{\min}
\right),
\end{equation}
where $\eta_{\min}$ denotes the minimum allowed learning rate, so that $\eta_t=\eta_e$ for every step $t$ belonging to epoch $e$. This strategy enables large parameter updates during the early stages of training while progressively refining the model as optimization approaches convergence. Such two-stage training protocols and adaptive learning-rate schedules have been employed in modern deep-learning-based interatomic potential frameworks to improve optimization stability and convergence~\cite{batatia2022mace, fu2025learning}.

A comparison of the optimization strategies used in the original NEP and TorchNEP frameworks is shown in Fig.~\ref{fig:workflow}d. The original NEP implementation updates model parameters using the SNES~\cite{schaul2011high}, while TorchNEP computes analytical gradients explicitly and performs mini-batch optimization using Adam. As a result, TorchNEP substantially reduces the computational cost of training while preserving the original NEP model formulation. 
% compatibility with NEP
To ensure compatibility with existing NEP workflows, the TorchNEP input format remains largely unchanged from the original NEP implementation, requiring only a few additional parameters associated with the PyTorch framework.
Table~\ref{tab:torchnep_input} shows a representative input file for the carbon dataset. 
All model-architecture parameters are inherited directly from the original NEP format, allowing existing NEP input files to be adapted with minimal modification. 
Additionally, consistent with the original NEP framework~\cite{liu2023largescale}, TorchNEP fully supports the Ziegler--Biersack--Littmark (ZBL) potential~\cite{ziegler1985treatise} as a short-range correction (Figure~S1). 

\begin{table}[htbp]
\centering
\caption{Example TorchNEP input file.}
\label{tab:torchnep_input}
\begin{tabular}{lll}
\hline
Category & Parameter & Value \\
\hline
\multirow{7}{*}{Model}
& type & 1 C \\
& version & 4 \\
& cutoff & 4.2 3.7 \\
& n\_max & 10 8 \\
& basis\_size & 10 8 \\
& l\_max & 4 2 1\\
& neuron & 100 \\
\hline
\multirow{8}{*}{Stage I}
& epoch & 600 \\
& lr & 5e-3 \\
& lambda\_e & 1.0 \\
& lambda\_f & 100.0 \\
& lambda\_v & 1.0 \\
& batch & 16 \\
& scheduler\_patience & 15 \\
& scheduler\_factor & 0.5 \\
\hline
\multirow{5}{*}{Stage II}
& stage2 & 1 \\
& start\_stage2 & 300 \\
& stage2\_lr & 5e-4 \\
& stage2\_lambda\_e & 100.0 \\
& stage2\_lambda\_f & 5.0 \\
& stage2\_lambda\_v & 30.0 \\
\hline
\end{tabular}
\end{table}

\subsection{Training efficiency and predictive accuracy across diverse atomistic datasets}

\begin{figure}[H]
    \centering
    \includegraphics[width=1\linewidth]{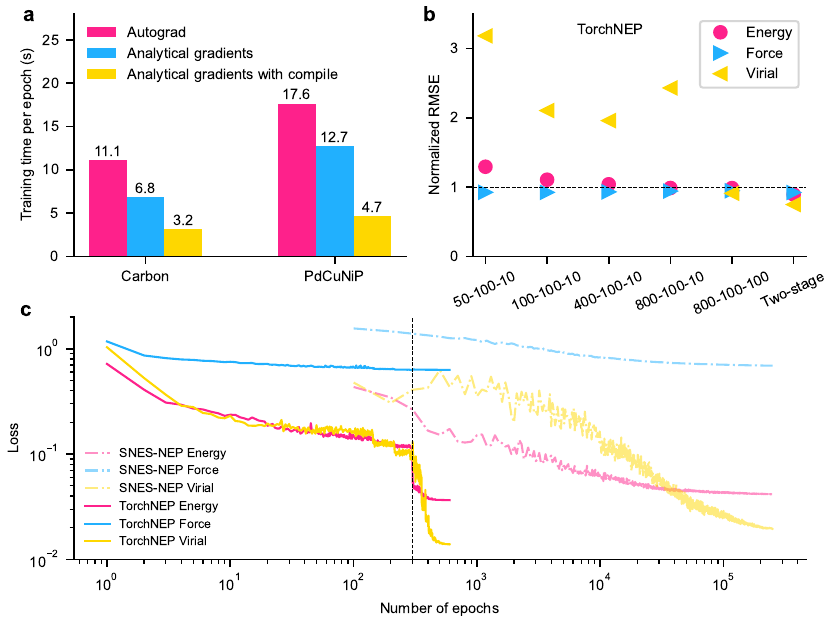}
    \caption{
\textbf{Effects of optimization algorithms and training strategies on efficiency and accuracy.}
(a) Training time per epoch on the carbon~\cite{carbon2017mechine,fan2022gpumd} and PdCuNiP alloy~\cite{zhao2023development} datasets using TorchNEP with autograd, analytically derived gradients, and the compiled analytical-gradient implementation.
(b) Normalized RMSEs relative to the published NEP model for TorchNEP models trained using different single-stage weight combinations and the two-stage training strategy on the carbon dataset. Labels such as 50–100–10 denote the weights assigned to energy, force, and virial losses, respectively.
(c) Evolution of energy, force, and virial training losses for SNES-NEP and TorchNEP on the carbon dataset. The black dashed line at epoch 300 marks the transition from Stage~1 to Stage~2 in the two-stage training protocol. All calculations were performed on a single NVIDIA V100 GPU.
    }
    \label{fig:sawhy}
\end{figure}

We first investigate the effects of optimization algorithms and training strategies on the efficiency and accuracy of TorchNEP. Figure~\ref{fig:sawhy}a compares the training time per epoch of TorchNEP using autograd and analytically derived force gradients on the carbon~\cite{carbon2017mechine} and PdCuNiP alloy~\cite{zhao2023development} datasets.
Several implementation-level optimizations accelerate both schemes. 
Neighbor lists, interatomic displacement vectors, and descriptor-related quantities are constructed once and retained on the GPU. 
With these optimizations, the autograd implementation requires approximately 11 s and 18 s per epoch for the carbon and PdCuNiP datasets, respectively.

The analytical-gradient implementation further improves efficiency by replacing automatic differentiation of forces with closed-form expressions derived from the descriptor--coordinate Jacobian and neural-network gradients. This eliminates the costly second-order backward pass required by autograd-based force training and enables additional optimizations, including precomputation and caching of geometry-dependent basis functions on the GPU. Consequently, the training time is reduced to 7 s and 13 s per epoch for the two datasets. Furthermore, the analytical formulation is compatible with \texttt{torch.compile}, which generates fused Triton GPU kernels and substantially reduces kernel-launch overhead and memory traffic. This optimization further decreases the training time to 3.2 s and 4.7 s per epoch. Overall, the analytical-gradient implementation achieves speedups of 3.4$\times$ and 3.8$\times$ relative to the already optimized autograd baseline.

We next investigate the origin of the accuracy improvements observed in TorchNEP.
Figure~\ref{fig:sawhy}b compares different training protocols on the carbon dataset. 
All published NEP models considered in this work employ a conventional single-stage training procedure with fixed loss weights throughout optimization. 
Increasing the energy weight from 50 to 800 while keeping the force and virial weights fixed at 100 and 10 reduces the energy RMSE from 52.9 to 40.4 meV/atom. Meanwhile, the force RMSE remains nearly unchanged, whereas the virial RMSE exhibits noticeable fluctuations due to its relatively small weight and the limited number of virial data in the training set. Increasing the virial weight further (800--100--100) improves the virial accuracy, yielding RMSEs of 40.3 meV/atom, 647.6 meV/\AA, and 17.0 meV for energy, force, and virial predictions, respectively, which are already comparable to those obtained using the original SNES training strategy ~\cite{fan2022gpumd} (40.9 meV/atom, 684.6 meV/\AA, and 18.7 meV). These results suggest that, under appropriately chosen loss weights, gradient-based and SNES-based optimization achieve a similar level of training accuracy. In contrast, the two-stage training protocol simultaneously improves all three metrics, achieving RMSEs of 36.5 meV/atom, 630.7 meV/\AA, and 13.9 meV. Therefore, the primary improvement in training accuracy originates from the training protocol rather than from the optimization algorithm itself.

The mechanism underlying this improvement is illustrated in Fig.~\ref{fig:sawhy}c, which compares the evolution of training losses for SNES-NEP and TorchNEP. During Stage~1, the large force weight rapidly reduces the force loss and enables the model to learn the overall shape of the potential-energy surface. In Stage~2, the energy and virial weights are increased while the force weight is reduced, allowing the model to refine local energetic details. As a result, both energy and virial losses exhibit a pronounced decrease after the transition to Stage~2, and the model reaches convergence within 600 epochs. By contrast, the SNES optimizer converges substantially more slowly and typically requires several hundred thousand optimization steps to achieve a comparable level of accuracy. Based on these observations, all subsequent TorchNEP models in this work employ analytical gradients together with graph-level compilation and the two-stage training protocol to achieve the best balance between efficiency and accuracy.

To assess the robustness of TorchNEP across different application domains, we benchmark it against carefully optimized published NEP models on a diverse collection of datasets, including carbon~\cite{carbon2017mechine,fan2022gpumd}, LiH~\cite{ying2025highly}, water~\cite{zhang2021phase,xu2023accurate}, Al/graphene (Al/Gra)~\cite{wu2025revealing}, CrCoNi~\cite{wu2026general}, PdCuNiP alloys~\cite{zhao2023development}, and BN/SiC/Cu interfaces~\cite{hashemi2026stabilization}. These datasets span a broad range of material classes, including elemental systems, molecular systems, metallic alloys, and heterogeneous interfaces, covering both solid and liquid phases. 
Unless otherwise specified, all TorchNEP models were trained for 600 epochs using the same model architecture as the corresponding published NEP models. For the original SNES-NEP implementation, training times were estimated from the first 100 epochs of the published input files, since the computational cost per epoch remains nearly constant during training. To ensure a fair comparison, only SNES-NEP runs originally configured with more than 300,000 epochs were truncated to 300,000 epochs, while all other runs retained their original settings.

\begin{figure}[H]
    \centering
    \includegraphics[width=1\linewidth]{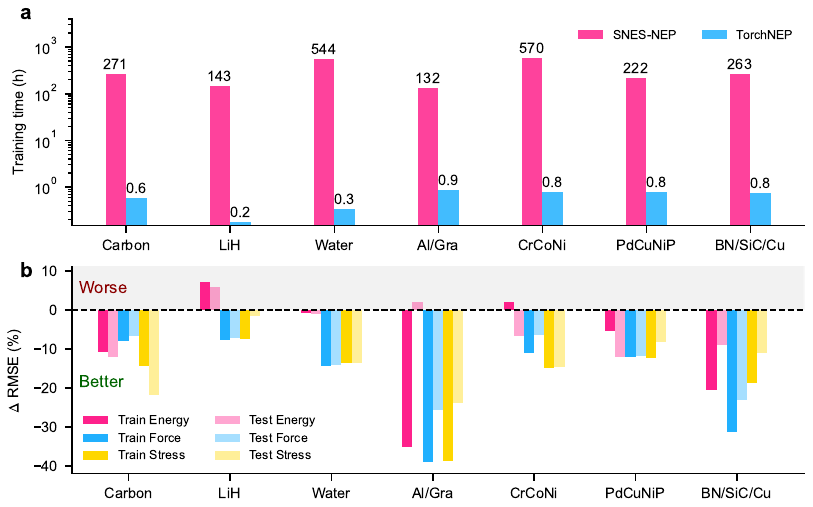}
    \caption{
    \textbf{Training efficiency and predictive accuracy across diverse datasets.}
    (a) Comparison of training time between the original NEP implementation and TorchNEP.
    (b) Relative changes in RMSE ($\Delta$RMSE) for energy, force, and stress obtained by TorchNEP with respect to the original NEP on both the training and test datasets.
    Benchmarks are performed on a diverse set of systems, including carbon~\cite{carbon2017mechine, fan2022gpumd}, LiH~\cite{ying2025highly}, water~\cite{zhang2021phase,xu2023accurate}, Al/graphene~\cite{wu2025revealing}, CrCoNi~\cite{wu2026general}, PdCuNiP alloys~\cite{zhao2023development}, and BN/SiC/Cu interfaces~\cite{hashemi2026stabilization}.
Positive values in (b) indicate higher RMSEs than those of the original NEP (worse predictive performance), whereas negative values indicate lower RMSEs (better predictive performance).}
    \label{fig:sa}
\end{figure}

As shown in Fig.~\ref{fig:sa}a, the combination of mini-batch training, gradient-based optimization, and compiler-assisted acceleration dramatically improves training efficiency compared with the SNES optimizer. On a single NVIDIA V100 GPU, TorchNEP reduces the training time from hundreds of hours to less than one hour for all tested datasets, corresponding to speedups of more than two orders of magnitude, with an average acceleration factor of approximately 628$\times$. The most striking example is the water dataset (1388 structures, $5.3 \times 10^{5}$ atoms), where the training time decreases from 544 hours for SNES-NEP to only 0.3 hours for TorchNEP.

We next compare the predictive accuracy of two-stage TorchNEP training and single-stage SNES-NEP training. Figure~\ref{fig:sa}b presents the relative changes in RMSE ($\Delta$RMSE) for energy, force, and stress with respect to the corresponding NEP models. Consistent with the observations in Fig.~\ref{fig:sawhy}, TorchNEP systematically improves force and stress accuracy on both the training and test datasets. The largest improvements are observed for the Al/graphene dataset, where the force and stress RMSEs are reduced by more than 30\% on the training set and more than 20\% on the test set.
For energy predictions, TorchNEP improves both training and test accuracy in five of the eight datasets. In the remaining cases, the differences are extremely small and unlikely to be practically significant. For example, the test-set energy RMSE of the Al/graphene model changes only from 59.0 to 60.0 meV/atom, while the training-set energy RMSE of the CrCoNi model changes from 2.45 to 2.50 meV/atom. The largest relative increase is observed for the LiH dataset; however, because the absolute errors are already extremely small, the change corresponds to only 0.096 to 0.103 meV/atom on the training set and 0.088 to 0.094 meV/atom on the test set. These variations are negligible compared with the overall accuracy of the models. The complete RMSE statistics are provided in Table~S1.

Overall, these benchmarks demonstrate that TorchNEP consistently delivers substantial gains in training efficiency while maintaining or improving predictive accuracy across a wide variety of material systems. Combined with the analysis in Fig.~\ref{fig:sawhy}, the results indicate that gradient-based optimization is primarily responsible for the dramatic acceleration in training, whereas the improved predictive accuracy mainly originates from the two-stage training protocol.
The two-stage training protocol is expected to improve the training accuracy of the SNES-NEP approach as well, but this has not been systematically explored yet.

\subsection{Parameter optimization and physical property prediction on 16 elemental systems}

While RMSE-based metrics provide a useful measure of fitting accuracy, they are often insufficient for assessing the transferability and physical fidelity of MLIPs, particularly for complex material systems. Moreover, understanding how training parameters and model complexity affect predictive accuracy and physical performance requires extensive benchmarking, which is readily enabled by the exceptional training efficiency of TorchNEP. To this end, we further evaluate TorchNEP on a comprehensive benchmark introduced by Song et al.~\cite{song2024general}, which contains 16 metallic elements (Ag, Al, Au, Cr, Cu, Mg, Mo, Ni, Pb, Pd, Pt, Ta, Ti, V, W, and Zr) and 105,464 DFT configurations with a total of 6.9 million atoms. The dataset covers a broad range of thermodynamic and mechanical conditions, with atomic forces ranging from -73 to 77 eV/\AA. 

Among the various hyperparameters in the NEP framework, the numbers of radial ($n_{\max}^{R}$) and angular ($n_{\max}^{A}$) descriptor components are key factors governing descriptor resolution and model expressivity. The original NEP model (denoted as NEPOri) trained by Song et al.~\cite{song2024general} employs $n_{\max}^{R}=4$ and $n_{\max}^{A}=4$. To investigate the influence of descriptor complexity on predictive accuracy and physical performance, three additional TorchNEP models with progressively increased descriptor sizes, namely NEP44, NEP66, and NEP88, were trained and benchmarked, as summarized in Table~\ref{tab:train_16elements}. Increasing both $n_{\max}^{R}$ and $n_{\max}^{A}$ from 4 to 8 substantially enlarges the model capacity, resulting in an increase in the number of trainable parameters from 70,401 to 124,673.
Nevertheless, all TorchNEP models complete training in approximately 2.8 hours on a single NVIDIA H200 GPU, whereas the original NEP implementation requires about 132 hours, corresponding to a more than 40-fold reduction in training time.
In terms of training accuracy, the training-set RMSEs decrease monotonically with increasing model size. 
Relative to NEPOri, NEP44 reduces the training RMSEs of energy, force, and stress by 12.3\%, 12.2\%, and 40.7\%, respectively. Notably, NEP44 employs exactly the same model architecture and number of trainable parameters as NEPOri. Therefore, these accuracy improvements can be attributed entirely to the training framework, in particular the two-stage training strategy discussed above. Further increasing the descriptor size yields additional reductions, with NEP88 achieving overall improvements of 30.8\%, 19.3\%, and 49.2\% for energy, force, and stress. These results indicate that larger models can substantially reduce training errors.

\begin{table}[H]
    \centering
    \begin{threeparttable}
    \caption{Training parameters and performance of different NEP models on the 16-element dataset.}
    \label{tab:train_16elements}
    \begin{tabular*}{\textwidth}{@{\extracolsep{\fill}}lcccccccc}
    \toprule
    Label &
    $n_{\max}^{R}$ &
    $n_{\max}^{A}$ &
    \makecell{Trainable \\ Parameter} &
    Implementation &
    \makecell{Training Time\tnote{a} \\(hours)} &
    \makecell{$E_{\mathrm{train}}$\\(meV/atom)} &
    \makecell{$F_{\mathrm{train}}$\\(meV/\AA)} &
    \makecell{$S_{\mathrm{train}}$\\(GPa)} \\
    \midrule
    NEPOri & 4 & 4 &  70,401 & SNES-NEP & 132 & 17.13 & 171.90 & 1.18 \\
    NEP44  & 4 & 4 & 70,401 & TorchNEP & 2.8 & 15.03 & 151.02 & 0.70 \\
    NEP66  & 6 & 6 & 97,537 & TorchNEP & 2.9 & 12.72 & 142.91 & 0.62 \\
    NEP88  & 8 & 8 & 124,673 & TorchNEP & 2.9  & 11.85 & 138.75 & 0.60 \\
    \bottomrule
    \end{tabular*}
        \begin{tablenotes}
        \footnotesize
        \item[a] The wall-clock time for NEP was estimated based on 300,000 generations, while that for TorchNEP was measured from actual 600-epoch training runs. All results were obtained on a single NVIDIA H200 GPU. 
        \end{tablenotes}
    \end{threeparttable}
\end{table}
 
\begin{figure}[H]
    \centering
    \includegraphics[width=1\linewidth]{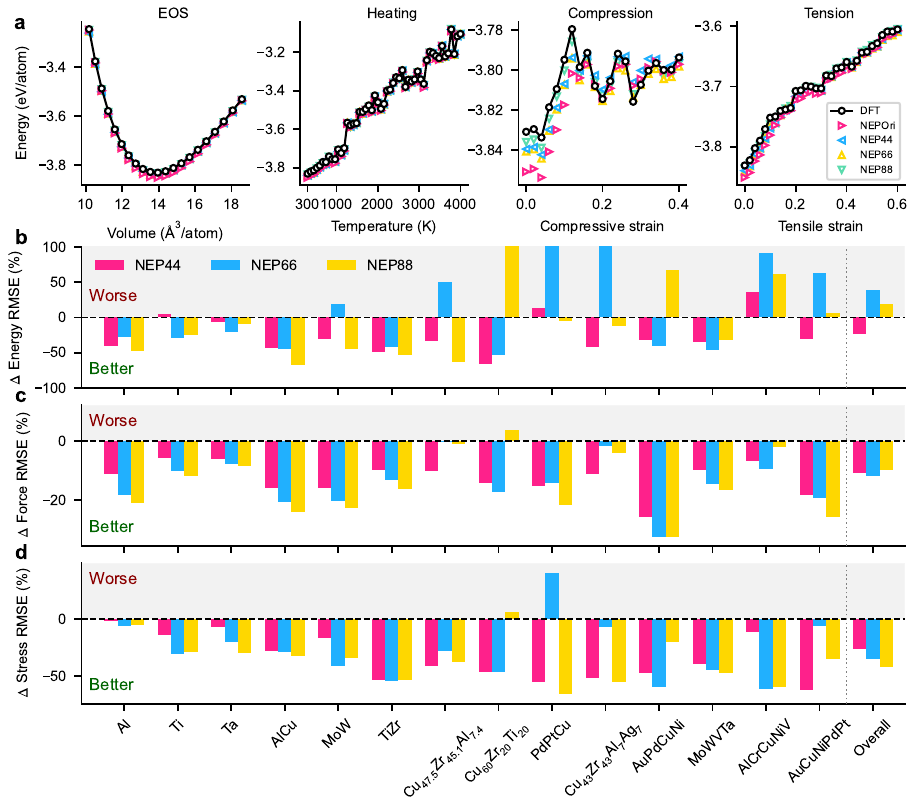}
    \caption{\textbf{Prediction performance on diverse test datasets.}
(a) Representative AlCu test dataset illustrating the benchmark protocol, including equation-of-state (EOS), heating, compression, and tension simulations. The heating, compression, and tension tests are included for all benchmark systems, while equation-of-state tests are included for alloy systems. The DFT reference values are taken from Ref.~\cite{song2024general}. Relative changes in RMSE ($\Delta$RMSE) with respect to NEPOri for (b) energy, (c) force, and (d) stress predictions across test datasets spanning elemental systems, binary alloys, metallic glasses, and high-entropy alloys. Negative values of $\Delta$RMSE indicate improved accuracy relative to NEPOri, whereas positive values indicate degraded performance.
}
    \label{fig:testset}
\end{figure}

We next evaluate the predictive performance of the four models on a diverse benchmark, as shown in Fig.~\ref{fig:testset}. 
Notably, while the training set contains only elemental and binary systems, the benchmark includes 14 subsets covering elemental systems, binary alloys, metallic glasses, and high-entropy alloys, thereby providing a stringent assessment of model transferability.
Figure~\ref{fig:testset}a takes the AlCu alloy as an example, comparing the predictions of the four models for the equation of state (EOS) and the equilibrium energies under varying temperatures, compressive strains, and tensile strains.
These configurations sample a broad range of thermodynamic and mechanical conditions, including heating from 300 to 4000 K, compression up to 40\% strain, tension up to 60\% strain, and static EOS calculations. Additional details of the benchmark construction can be found in Ref.~\cite{song2024general}.

The test-set results reveal a more nuanced trend than that observed for the training errors. As shown in Fig.~\ref{fig:testset}b--d, NEP44 consistently improves force and stress predictions across all benchmark systems, reducing the overall force and stress RMSEs by approximately 10.7\% and 26.0\%, respectively, relative to NEPOri. NEP66 further reduces the overall force and stress RMSEs; however, the improvement is not uniform across all subsets. For example, its stress prediction for the PdPtCu system is worse than that of NEPOri (Fig.~\ref{fig:testset}d). NEP88 achieves the lowest overall stress RMSE, but its force RMSE is slightly larger than those of NEP44 and NEP66.

The behavior of the energy predictions differs from that of force and stress. While NEP44 reduces the overall energy RMSE by 23.5\% relative to NEPOri, the overall energy RMSE increases by 39.3\% and 19.1\% for NEP66 and NEP88, respectively. At the level of individual systems, the variation is even more pronounced. For example, NEP66 exhibits substantially larger energy errors for PdPtCu and Cu$_{43}$Zr$_{43}$Al$_7$Ag$_7$, while NEP88 produces a significantly larger error for Cu$_{60}$Zr$_{20}$Ti$_{20}$. These results indicate that reductions in training error do not necessarily translate into uniform improvements across unseen systems. The complete test-set RMSE statistics are provided in Table~S2.

\begin{figure}[H]
    \centering
    \includegraphics[width=1\linewidth]{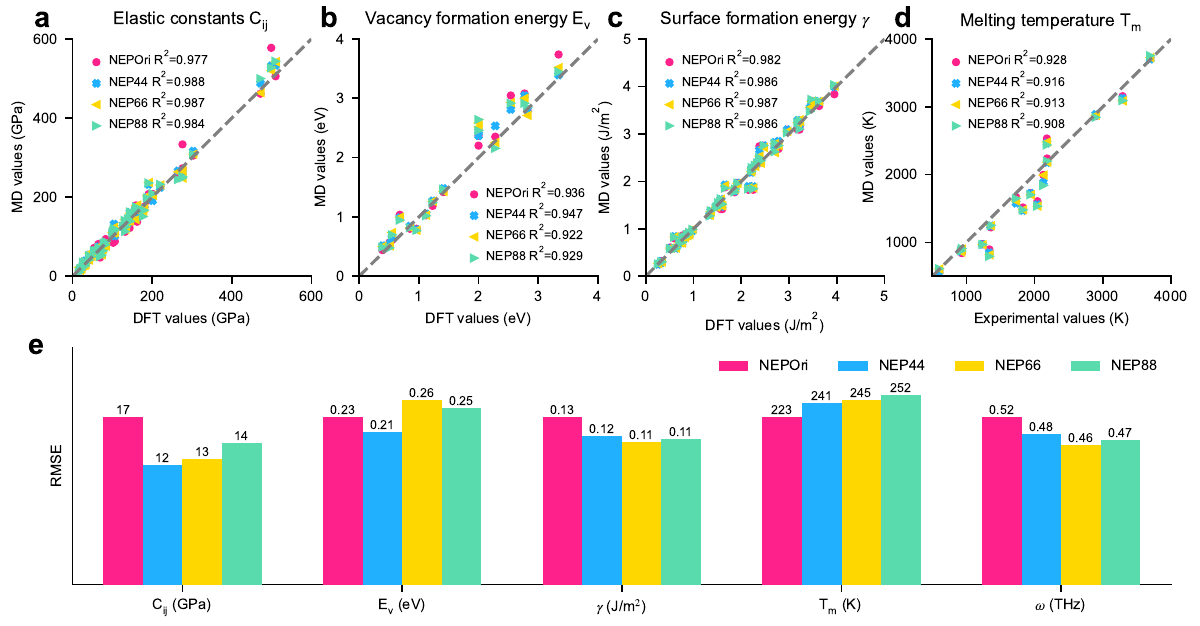}
    \caption{\textbf{Predictive performance on physical material properties.}
Parity plots comparing reference values and MD-predicted values for (a) elastic constants $C_{ij}$, (b) vacancy formation energies $E_v$, (c) surface energies $\gamma$, and (d) melting temperatures $T_m$ across 16 elemental systems (Ag, Al, Au, Cr, Cu, Mg, Mo, Ni, Pb, Pd, Pt, Ta, Ti, V, W, and Zr). Reference values for $C_{ij}$, $E_v$, and $\gamma$ are obtained from DFT calculations\cite{de2015charting, tran2016surface, song2024general}, whereas $T_m$ is compared against experimental measurements~\cite{haynes2016crc}. Corresponding $R^2$ (coefficient of determination) values are provided in each panel. (e) Comparison of RMSE values for $C_{ij}$, $E_v$, $\gamma$, $T_m$, and phonon frequencies $\omega$ obtained using the four models. For each property, bar heights are normalized by the corresponding RMSE of NEPOri, while the numerical labels indicate the absolute RMSE values. Complete phonon dispersion comparisons for all elemental systems are provided in Fig.~S2--S4.}
    \label{fig:phypro}
\end{figure}

To further evaluate whether the trends observed in the configuration-level benchmark translate into practical materials predictions, we compare the four models on a diverse set of physical properties derived from molecular dynamics simulations. Figure~\ref{fig:phypro} summarizes the prediction accuracy for elastic constants ($C_{ij}$), vacancy formation energies ($E_v$), surface energies ($\gamma$), melting temperatures ($T_m$), and phonon frequencies ($\omega$) across the 16 elemental systems. Figures~\ref{fig:phypro}a--d show parity plots for $C_{ij}$, $E_v$, $\gamma$, and $T_m$. All four models reproduce the overall trends well, yielding coefficients of determination ($R^2$) greater than 0.9 for all properties considered.

To facilitate quantitative comparison, Fig.~\ref{fig:phypro}e summarizes the RMSE values of all five properties. Consistent with the configuration-level benchmark shown in Fig.~\ref{fig:testset}, NEP44 systematically improves the prediction of all DFT-derived properties relative to NEPOri. The RMSE of elastic constants decreases from 17.1 to 12.1 GPa, while the RMSEs of vacancy formation energies, surface energies, and phonon frequencies decrease from 0.233 to 0.212 eV, from 0.131 to 0.116 J/m$^2$, and from 0.523 to 0.485 THz, respectively. Detailed DFT reference values for $C_{ij}$, $E_v$, and $\gamma$~\cite{de2015charting, tran2016surface, song2024general}, together with the corresponding predictions of the four NEP models for all 16 elements, are provided in Table~S3--S6. Complete phonon dispersion comparisons are also presented in Fig.~S2--S4. These results indicate that the accuracy gains introduced by the two-stage training strategy are not limited to fitting metrics but also translate into improved predictive performance for practical physical properties.

Further increasing the descriptor size produces mixed outcomes. NEP66 achieves the lowest RMSEs for surface energies and phonon frequencies, suggesting that larger descriptors can improve certain energetic and vibrational properties. However, the accuracy of elastic constants deteriorates relative to NEP44, and the vacancy formation energy RMSE increases to 0.26 eV, exceeding even that of NEPOri. Similar behavior is observed for NEP88, which further improves the prediction of phonon frequencies and surface energies but exhibits worse performance for both elastic constants and vacancy formation energies compared with NEP44. These observations indicate that improvements in individual properties do not necessarily translate into systematic improvements across all physical properties.

A different trend is observed for melting temperatures. The RMSE increases moderately from 223 K for NEPOri to 241 K, 245 K, and 252 K for NEP44, NEP66, and NEP88, respectively. The corresponding experimental reference values and model predictions are summarized in Table~S7. Unlike the other properties considered here, melting temperatures are compared against experimental measurements rather than DFT calculations. Consequently, these variations should not be interpreted as a direct measure of the intrinsic accuracy of the potentials and are likely influenced by both experimental uncertainty and approximations in the atomistic simulations.

Combining the results of Figs.~\ref{fig:testset} and \ref{fig:phypro}, a consistent picture emerges. The accuracy improvements introduced by the two-stage training strategy are reflected not only in lower training errors but also in improved predictions for unseen configurations and DFT-derived physical properties. By contrast, although increasing descriptor size continuously reduces the training errors, the resulting gains do not systematically transfer to test-set performance or physical-property predictions. Among the models considered here, NEP44 provides the most balanced performance across training accuracy, predictive accuracy, and physical-property evaluation.

\section{Conclusion}

In this work, we developed TorchNEP, a PyTorch-based implementation of the neuroevolution potential (NEP) framework. By combining analytical gradients, mini-batch training, adaptive optimization, and compiler-assisted acceleration, TorchNEP maintains full compatibility with existing NEP models while reducing training times by more than two orders of magnitude compared with the original SNES-based implementation. Beyond efficiency, we systematically analyzed the origin of the accuracy improvements achieved by TorchNEP. Our results show that gradient-based optimization alone yields training accuracy comparable to that of the original NEP implementation, whereas the primary accuracy gain originates from the two-stage training strategy. Across a diverse collection of benchmark datasets, TorchNEP consistently improves force and stress predictions while maintaining comparable or improved energy accuracy. The significantly improved training efficiency of TorchNEP further enables practical exploration of model complexity. Using a comprehensive 16-element benchmark, we show that larger descriptor sizes can further reduce training errors but do not necessarily improve predictive performance on unseen configurations and physical properties. These results highlight the importance of evaluating machine-learned interatomic potentials beyond training metrics alone. Overall, TorchNEP provides an efficient and flexible training framework for NEP models and makes systematic hyperparameter exploration computationally practical. We expect it to facilitate the development of more accurate and transferable machine-learned interatomic potentials for large-scale atomistic simulations.

\section{Methods}

\subsubsection{NEP descriptors}

The NEP descriptor vector $\bm{q}^{i}$ consists of radial (two-body) and angular
(three- and higher-body) components. Two independent cutoff radii are used:
the radial cutoff $r_c^{\mathrm{R}}$ and the angular cutoff
$r_c^{\mathrm{A}}$. For each radial channel
$n=0,\dots,n_{\max}^{\mathrm{R}}$,
the radial descriptor is defined as
\begin{equation}
  q^{i}_{n}
  =
  \sum_{j\neq i}
  g_{n}(r_{ij}),
\end{equation}
where
\begin{equation}
  g_{n}(r_{ij})
  =
  \sum_{k=0}^{N_{\mathrm{bas}}^{\mathrm{R}}}
  c^{\,t_i t_j}_{nk}\,
  f_{k}(r_{ij}).
\end{equation}
Here,
$r_{ij}=|\bm{r}_j-\bm{r}_i|$,
and
$c^{\,t_i t_j}_{nk}$
are trainable coefficients associated with the species pair
$(t_i,t_j)$. The basis functions are constructed from Chebyshev polynomials
multiplied by a smooth cutoff function,
\begin{equation}
  f_{k}(r)
  =
  \frac{1}{2}
  \left(T_k\left[2
  \left(
  \frac{r}{r_c}-1
  \right)^2
  -1\right]+1\right)
  f_c(r),
\end{equation}
where the cutoff function is defined as
\begin{equation}
  f_c(r)
  =
  \begin{cases}
    \frac{1}{2}
    \left[
      \cos\!\left(\pi r/r_c\right)+1
    \right],
    & r \le r_c,
    \\[3pt]
    0,
    & r > r_c.
  \end{cases}
\end{equation}
Both $f_k(r)$ and its first derivative vanish continuously at
$r=r_c$.

Angular information is represented through expansion moments of the neighbor
density,
\begin{equation}
  A^{i}_{nlm}
  =
  \sum_{j\neq i}
  g^{\mathrm{A}}_{n}(r_{ij})
  Y_{lm}(\hat{\bm{r}}_{ij}),
\end{equation}
where $\hat{\bm{r}}_{ij}$ is the unit bond vector,
$Y_{lm}$ denotes the  spherical harmonic of angular order
$l=1,\dots,l_{\max}$,
and
$g^{\mathrm{A}}_{n}$
adopts the same Chebyshev expansion form as the radial basis but uses the
angular cutoff and an independent set of trainable coefficients. The three-body descriptor is obtained from the rotationally invariant
quadratic contraction
\begin{equation}
  q^{i}_{nl}
  =
  \sum_{m=-l}^{l}
  \left|A^{i}_{nlm}\right|^2.
  \label{eq:q3b}
\end{equation}

Higher-body descriptors are constructed from higher-order contractions of the
expansion moments, and their explicit forms can be found in
Refs.~\cite{fan2022gpumd}. By combining all radial, three-body, and higher-body components,
the complete descriptor vector
$\bm{q}^{i}\in\mathbb{R}^{N_{\mathrm{des}}}$
is obtained. By construction, the NEP descriptor is invariant under translation, rotation,
and permutation of identical atoms. Translational invariance follows because the
descriptor depends on the atomic positions only through the relative vectors
$\mathbf{r}_{ij} = \mathbf{r}_j - \mathbf{r}_i$. Rotational invariance holds
because the radial part depends only on the distances $d_{ij} =
\lVert \mathbf{r}_{ij} \rVert$, while the angular part is built from rotationally
invariant contractions of the bond directions. Permutational invariance arises
because each component sums over neighbors within the cutoff, which is
independent of their ordering. Consequently, the predicted energy is invariant,
and the derived forces and virials transform covariantly, under these
operations.

\subsubsection{Evaluations of basic physical properties}

To assess the transferability of the trained models beyond fitting errors, we evaluated a series of static and dynamic material properties, including elastic constants, vacancy formation energies, surface energies, melting temperatures, and phonon dispersions, as summarized in Fig.~\ref{fig:phypro}. Elastic constants were obtained using the stress--strain method. Starting from fully relaxed equilibrium structures, small Lagrangian strains were applied along independent deformation modes, followed by relaxation of the internal atomic coordinates. The resulting Cauchy stress tensor was evaluated and the elastic constants were extracted from the linear stress--strain relationship.

Melting temperatures of the 16 elemental systems were determined using the solid--liquid coexistence method. For each element, a two-phase supercell with a length of approximately 70~\AA\ along the $x$ direction and 36~\AA\ along the remaining directions was constructed. One half of the simulation cell was retained in the crystalline state, while the other half was melted using a Langevin thermostat at $2T_{\mathrm{guess}}$, where $T_{\mathrm{guess}}$ denotes the current estimate of the melting temperature. After generating a stable solid--liquid interface, the system was evolved in the isobaric--isoenthalpic (NPH) ensemble at zero pressure for 400 ps. The evolution of the interface was then used to determine whether the trial temperature was above or below the melting point. If the liquid region gradually solidified, $T_{\mathrm{guess}}$ was increased; conversely, if the solid phase melted completely, $T_{\mathrm{guess}}$ was decreased. This procedure was repeated iteratively until a stable coexistence between the solid and liquid phases was maintained throughout the simulation. The melting temperature was subsequently obtained by averaging the instantaneous temperature over the final 100 ps of the coexistence simulation . 

All structure generation, property calculations, and post-processing analyses were performed using the MDAPY package~\cite{mdapy}. Phonon dispersions were calculated through the PHONOPY package~\cite{phonopy} using the MDAPY~\cite{mdapy} interface. All melting simulations were carried out using LAMMPS~\cite{lammps}.

\section{Data availability}
All trained models and input files used in this work will be made publicly available upon acceptance of the manuscript. The training and test datasets are publicly available from previously published studies.

\section{Code availability}
The source code for MDAPY is available at the GitHub repository: https://github.com/mushroomfire/mdapy.

The source code for LAMMPS is available at the GitHub repository: https://github.com/lammps/lammps.

The source code for GPUMD is available at the GitHub repository: https://github.com/brucefan1983/GPUMD. 

The source code for TorchNEP will be made publicly available upon acceptance of the manuscript. 

\section{Author Contributions Statement}

Y.-C.W. developed the code, performed the simulations, analyzed the data, visualized the results, and contributed to writing the manuscript. X.C. contributed to data visualization and manuscript preparation. T.M., A.E., J.-L.S. and T.A.-N. contributed to results analysis and manuscript preparation. Z.F. contributed to the theoretical framework, data analysis and manuscript preparation. M.A. conceived and supervised the project and contributed to manuscript preparation. All authors reviewed and approved the final manuscript.

\begin{acknowledgments}

Y.-C.~W., T.~M. and M.~A. acknowledge the support from FinnCERES flagship (grant no.~151830423), Business Finland (grant nos.~211835, 211909, and 211989), the Research Council of Finland (grant no.~13361245), and the Future Makers program. 
M.~A. acknowledges support from the Academy of Finland Center of Excellence program (program nos. 278367 and 317464), as well as the Finnish Cultural Foundation.
A.~E. acknowledges support from the European Union Horizon 2020 research and innovation program under Grant Agreement No.~857470, from the European Regional Development Fund under the Foundation for Polish Science International Research Agenda PLUS program (Grant No.~MAB PLUS/2018/8), and from the initiative of the Ministry of Science and Higher Education ``Support for the activities of Centers of Excellence established in Poland under the Horizon 2020 program'' (Agreement No.~MEiN/2023/DIR/3795).
T.A-N. and X. C. have been supported in part by the Academy of Finland grants nos. 370057 and 373647. 
Z.F was supported by the Advanced Materials-National Science and Technology Major Project (No. 2024ZD0606900).
The authors acknowledge the computational resources provided by the Aalto University School of Science ``Science-IT'' project, as well as by CSC (Finland) \textit{via} the project 2015437.

\end{acknowledgments}

\bibliography{myref}

\end{document}